\documentclass[preprint,aps,showpacs]{revtex4}
\usepackage{graphicx}
\usepackage{amsmath}
\usepackage{amssymb}
\usepackage{hyperref}
\usepackage{latexsym}
\usepackage{epsfig}

\begin{document}

\title{Controlling the quantum stereodynamics of ultracold bimolecular reactions}

\author{M. H. G. de Miranda,$^{1\ast}$ A. Chotia,$^{1\ast}$ B. Neyenhuis,$^{1\ast}$ D. Wang,$^{1\ast}$$^{\S}$ G. Qu\'em\'ener,$^{1}$ S. Ospelkaus,$^{1}$ J. L. Bohn,$^{1}$ J. Ye,$^{1\dagger}$ D. S. Jin$^{1\dagger}$}
\affiliation{
\normalsize{$^{1}$JILA, NIST and University of Colorado,}\\
\normalsize{Department of Physics~University of Colorado}\\
\normalsize{Boulder, CO 80309-0440, USA}\\
\normalsize{$^\ast$These authors contributed equally to this work; }\\
\normalsize{$^\S$Present address: Department of Physics, The Chinese University of Hong Kong; }\\
\normalsize{$^\dagger$To whom correspondence should be addressed; }\\
\normalsize{E-mail:  Ye@jila.colorado.edu; Jin@jilau1.colorado.edu}
}

\pacs{XXX}

\maketitle

\textbf{Chemical reaction rates often depend strongly on
stereodynamics, namely the orientation and movement of molecules in
three-dimensional space~\cite{Zare,Aldegunde05,Taatjes07}. An
ultracold molecular gas, with a temperature below 1 $\mu$K, provides
a highly unusual regime for chemistry, where polar molecules can
easily be oriented using an external electric field and where,
moreover, the motion of two colliding molecules is strictly
quantized. Recently, atom-exchange reactions were observed in a
trapped ultracold gas of KRb molecules~\cite{OspelScience}. In an
external electric field, these exothermic and barrierless
bimolecular reactions, KRb+KRb$\rightarrow$ K$_{2}$+Rb$_{2}$, occur
at a rate that rises steeply with increasing dipole
moment~\cite{NiNature}. Here we show that the quantum stereodynamics
of the ultracold collisions can be exploited to suppress the
bimolecular chemical reaction rate by nearly two orders of magnitude. We use an
optical lattice trap to confine the fermionic polar molecules in a
quasi-two-dimensional, pancake-like geometry, with the dipoles
oriented along the tight confinement
direction~\cite{Buchler2D,Micheli2D}. With the combination of
sufficiently tight confinement and Fermi statistics of the
molecules, two polar molecules can approach each other only in a
``side-by-side'' collision, where the chemical reaction rate is
suppressed by the repulsive dipole-dipole interaction. We show that
the suppression of the bimolecular reaction rate requires
quantum-state control of both the internal and external degrees of
freedom of the molecules. The suppression of chemical reactions for
polar molecules in a quasi-two-dimensional trap opens the way for
investigation of a dipolar molecular quantum gas. Because of the
strong, long-range character of the dipole-dipole interactions, such
a gas brings fundamentally new abilities to quantum-gas-based
studies of strongly correlated many-body physics, where quantum
phase transitions and new states of matter can
emerge~\cite{Micheli,Baranov,Pfau,Demler,Santos,Capo}.}


Two colliding polar molecules interact via long-range dipole-dipole
forces well before they reach the shorter distance scales where
chemical forces become relevant.  Therefore, the spatial anisotropy
of the dipolar interaction can play an essential role in the
stereochemistry of bimolecular reactions of polar molecules. In
general, one expects the attraction between oriented dipoles in a
``head-to-tail'' collision to be favorable for chemical reactions,
while the repulsion between two oriented polar molecules in a
``side-by-side" collision presents an obstacle for reactions. Up to
now, however, large center-of-mass collision energies have precluded
the direct control of chemical reactions via dipolar interactions.
In a cold collision regime, where tens of scattering partial waves
contribute, one can begin to exert control of intermolecular
dynamics through the dipolar effect~\cite{Sawyer10}. An ultracold
gas, however, provides an optimum environment in which to fully
investigate the dipolar effects~\cite{NiNature,Krems,reviewpolar}.
Here, the molecules can be prepared in identical internal quantum
states, with the dipoles oriented using an external electric field, and
the molecular gas confined in external potentials created using
light. In the limit of vanishing collision energies, the
stereodynamics is described by only a few quantized collision
channels, and, moreover, for indistinguishable molecules, the states
of translational motion are coupled to internal molecular states due
to the fact that the quantum statistics of the molecules (fermions
or bosons) dictates a particular symmetry of the total wavefunction
with respect to exchange of two molecules.  In this quantum regime,
we have an opportunity to suppress or enhance reaction rates by
understanding and precisely controlling the stereodynamics of
colliding polar molecules.

\begin{figure}[htbp]
        \centering
                     \includegraphics[width=1.00\textwidth]{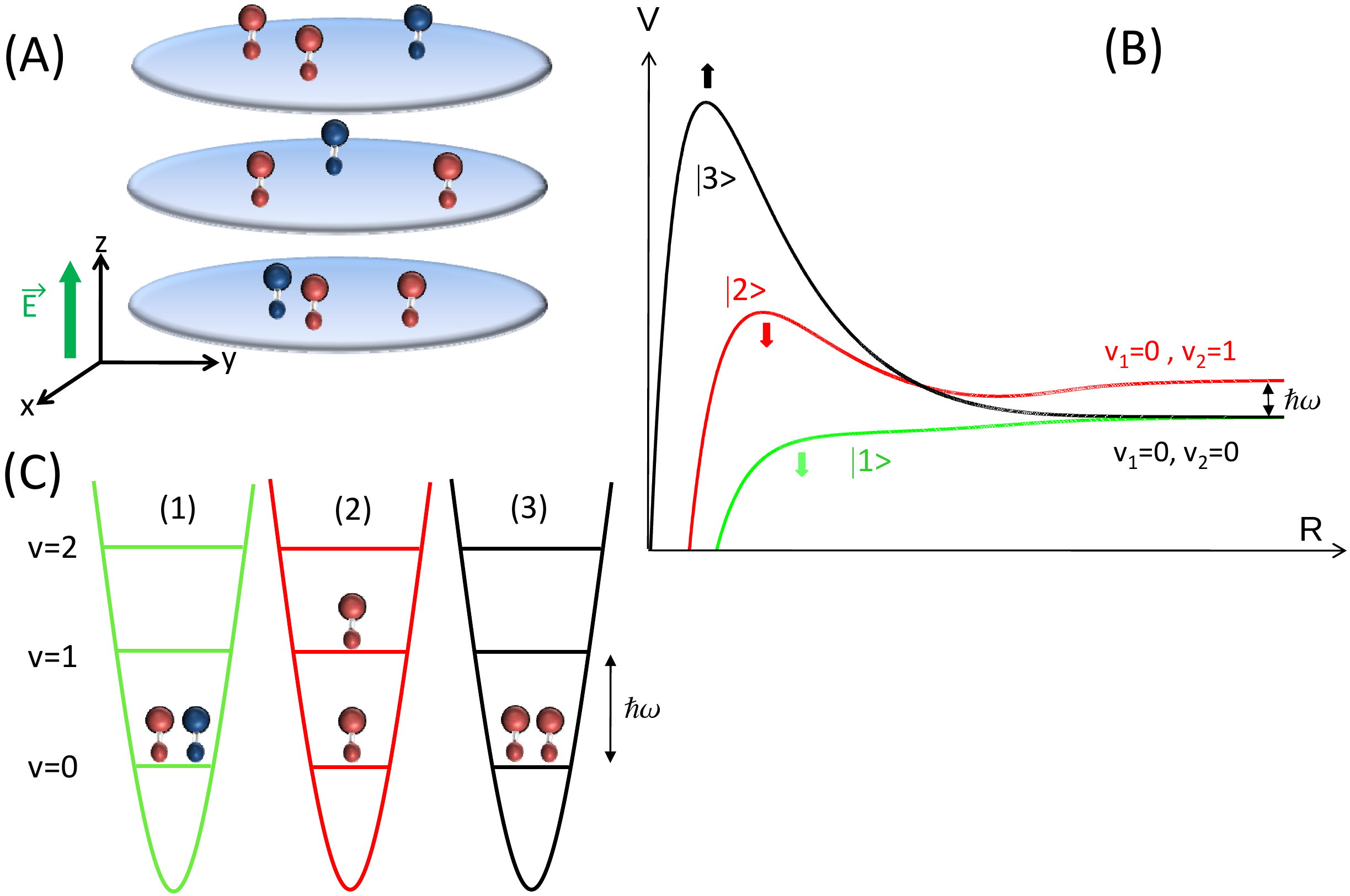}
                \caption{\textbf{Quantized stereodynamics of ultracold chemical reactions in quasi-2D.} (A)
                A quasi-2D geometry for collisions is realized for polar molecules confined in a one-dimensional
                optical lattice. An external electric field is applied along the tight confinement axis.
                (B) The lowest three adiabatic potentials for collisions are shown schematically
                as a function of the intermolecular separation, $R$. These three channels are ordered with increasing magnitude of the centrifugal
                barrier. The arrows indicate the change in the potential for an increasing external electric field, and hence a growing induced
                dipole moment. (C) Three different cases are shown schematically for each of the three
                lowest collision channels. The lowest energy collision channel occurs when two molecules are
                prepared in different internal states (indicated here by the colors of the molecules).
                The second channel is realized when two identical molecules are prepared in different
                vibrational levels $v$ for their $\hat{z}$ motions. The third case has a much reduced
                loss rate as a consequence of an increased centrifugal barrier when the two identical molecules are prepared in the same vibrational level along $\hat{z}$. }

            \label{fig1}
        \end{figure}

The spatial geometry of the confining potential can influence
collisions in a trapped gas of polar molecules.  In particular, a
two-dimensional (2D) trap geometry, with the dipoles oriented
parallel to the tight confinement direction $\hat{z}$, is
well-matched to the spatial anisotropy of the dipole-dipole
interaction~\cite{Tickner,Quemener2,Micheli3}.  We can realize such
a geometry using a one-dimensional optical lattice (see
Fig.~\ref{fig1}~A), where the trapped molecules are divided among
several isolated layers.  In each of these layers, the lattice
potential provides tight harmonic confinement in $\hat{z}$ such that
only the lowest few quantized motional states in $\hat{z}$ are
occupied. Consequently, within each isolated layer, colliding
molecules approach each other in 2D. However, the range of the van
der Waals interaction (and, for that matter, the range of dipolar
interactions at our largest external electric field) is still
smaller than the spatial extent of the cloud in the direction of
tight confinement, $a_{ho}$, and, therefore, at short intermolecular
distances a collision still must be treated in three dimensions
(3D).

We now consider the quantized collision channels that define the
stereodynamics in this quasi-2D geometry. For intermolecular
separations that are much larger than $a_{ho}$, the relative motion
of the two molecules is described by a quantized angular momentum,
$\hbar M$, around the $\hat{z}$ axis, where $\hbar=\frac{h}{2\pi}$
and $h$ is Planck's constant, as well as by a quantized relative
motion along $\hat{z}$. As discussed above, the stereodynamics of
ultracold collisions of indistinguishable molecules is strongly
influenced by the fact that the two-molecule wavefunction must obey
an overall symmetry with respect to the exchange of the identical
molecules. Therefore, an essential aspect of
relative motion in $\hat{z}$ is the exchange symmetry of this part
of the two-molecule wavefunction, which we identify with a quantum
number $\gamma$; for the symmetric case, $\gamma=1$ and for the
antisymmetric case, $\gamma=-1$. For two molecules in the same
$\hat{z}$ harmonic oscillator state, $\gamma=1$; while both
$\gamma=1$ and $\gamma=-1$ are possible for two molecules in
different harmonic oscillator states. Similarly, we use a quantum
number $\eta$ to keep track of the exchange symmetry of the part of
the wavefunction that describes the internal states of the two
molecules.  For two molecules in the same internal quantum state,
$\eta=1$, while $\eta=\pm1$ for molecules in different internal
states.  In 2D, these three quantum numbers ($M, \gamma, \eta$) are
sufficient to describe the quantum stereodynamics. However, because
the interactions at short range must be described in 3D, the quantum
number corresponding to the 3D angular momentum, $L$, as well as
$M$, becomes relevant. With collisional channels described by
quantum numbers $\eta, L, \gamma,$ and $M$, the fermionic symmetry
can be concisely stated in the following relations:
\begin{eqnarray}
    \eta(-1)^{L}=-1, ~~~~~~~~~~~\text{Short range, 3D}
    \label{SR}
    \\ \eta\gamma(-1)^{M}=-1.~~~~~~~~~~\text{Long range, 2D}
    \label{LG}
\end{eqnarray}

For ultracold collisions, the chemical reaction rate will be
dominated by the allowed collision channel with the lowest
centrifugal barrier. Combining this fact with the relations above,
we identify three collision channels relevant to the stereodynamics,
and we label these $|1\rangle$, $|2\rangle$, and $|3\rangle$, in
order of increasing centrifugal barrier heights.  The dipole-dipole
interaction mixes states with different $L$. However, for
convenience, we will refer to the lowest energy adiabatic channel,
which does not have a centrifugal barrier, as $L=0$. Similarly, we
will use $L=1$ to denote the odd-$L$ adiabatic channel with the
lowest centrifugal barrier. Collision channel $|1\rangle$ has
$\eta=-1$, $L=0$, $\gamma=1$, and $M=0$, and corresponds to
spatially isotropic collisions. Collision channel $|2\rangle$ has
$\eta=1$, $L=1$, $\gamma=-1$, and $M=0$, and is the quantum analog
of ``head-to-tail'' collisions. Collision channel $|3\rangle$ has
$\eta=1$, $L=1$, $\gamma=1$, and $M=\pm1$, and is the quantum analog
of ``side-by-side'' collisions. Fig.~\ref{fig1}~B shows
schematically the adiabatic potentials for these three lowest energy
collision channels.  Channels $|1\rangle$ and $|2\rangle$ become
increasingly favorable for chemical reactions as the dipole-dipole
interaction strength is increased, for example by increasing the
external electric field $\overrightarrow{E}$.  In contrast, channel
$|3\rangle$ has a centrifugal barrier whose height increases for
higher dipole moment, within the $|\overrightarrow{E}|$ range
considered in this work. This barrier hence continues to prevent
molecules from reaching short range.

Fig.~\ref{fig1}~C shows how these different collision channels can
be accessed through control of the internal molecular states and the
$\hat{z}$ motional states. In Fig.~\ref{fig1}~C, molecules in
different internal states are shown in different colors and the
harmonic oscillator states in $\hat{z}$ are labeled by $v$. In case
(1), for two molecules in different internal molecular states and in
any combination of $v$ levels, channel $|1\rangle$ is allowed when $\eta=-1$, resulting in no
centrifugal barrier.  In case (2), when
the molecules are prepared in identical internal molecular states
but in different $v$ levels, the lowest energy collision channel is
$|2\rangle$ (``head-to-tail''), which is allowed when $\gamma=-1$.
In case (3), where the molecules are prepared in the same internal
state and the same $v$ level, the two lower energy collision
channels are no longer allowed, and reactions can only proceed
through channel $|3\rangle$ (``side-by-side''). This case is
the least favorable for atom-exchange bimolecular chemical reactions.

We create a trapped, ultracold gas of $^{40}$K$^{87}$Rb molecules,
in their lowest energy ro-vibrational level and in a single
hyperfine state~\cite{hyperfine}, following the techniques described
in Ref.~\cite{NiScience}. To confine the molecules, we start with a
crossed-beam optical dipole trap, with a harmonic trapping frequency
of 180 Hz along the vertical direction ($\hat{z}$) and 25 Hz in the
transverse directions. For the current work, we add an optical
lattice along $\hat{z}$, which is formed by a retro-reflected beam
with a $1/e^{2}$ waist of $250~\mu$m and a wavelength of $1064$ nm.
Both optical dipole trap beams and the optical lattice beam are
linearly polarized and their polarizations are mutually orthogonal.
Each layer of the optical lattice trap is tightly confining in
$\hat{z}$ with a harmonic trapping frequency of $\nu_z=$ 23 kHz for the molecules,
while in the transverse directions, the combined trap has a harmonic
trapping frequency of 36 Hz. The tunneling rate between lattice
layers is negligible and, therefore, each layer realizes an isolated
trap for the molecules. Initially,
34,000 ground-state molecules are confined in roughly 23 layers,
with the center layer having 2200 molecules and a peak density of
3.4$\times10^{7}$ cm$^{-2}$.

We start by loading ultracold $^{40}$K and $^{87}$Rb atoms from the
crossed-beam dipole trap into the combined trap by turning up the
intensity of the optical lattice beam in 150 ms. We then create
molecules in the lattice by first forming extremely weakly bound
molecules through magneto-association of atom pairs and then
coherently transferring these molecules into their ro-vibrational
ground state using optical transitions~\cite{NiScience}. The
temperature of the molecular gas, $T$, in the combined optical
dipole plus lattice trap can be varied between 500 nK and 800 nK by
varying the initial atom gas conditions.  To completely freeze out
motion of the molecules along $\hat{z}$ requires that $k_B T \ll h
\nu_z$, where $k_B$ is Boltzmann's constant.  For a gas at $T=800$
nK in our lattice, $\frac{k_B T}{h \nu_z}=$ 0.72 , and we expect
$25\%$ of the molecules will occupy higher $v$ levels.

As discussed above, in order to control the stereochemistry of bimolecular reactions in the ultracold gas, we need to control both the internal state and the harmonic oscillator level $v$ of the molecules. We create the molecules in a single internal quantum state. If desired, we can subsequently create a 50/50 mixture of molecules in the ground and first excited rotational states by applying a resonant microwave $\pi/2$-pulse~\cite{hyperfine}.  The occupation of lattice levels $v$ can be controlled by varying $T$; alternatively, we can prepare a non-thermal distribution of molecules using parametric heating. Here, the lattice intensity is modulated at twice $\nu_{z}$, and, as a result, molecules initially in the $v=0$ level are excited to the $v=2$ level.

\begin{figure}[htbp]
        \centering
                      \includegraphics[width=1.00\textwidth]{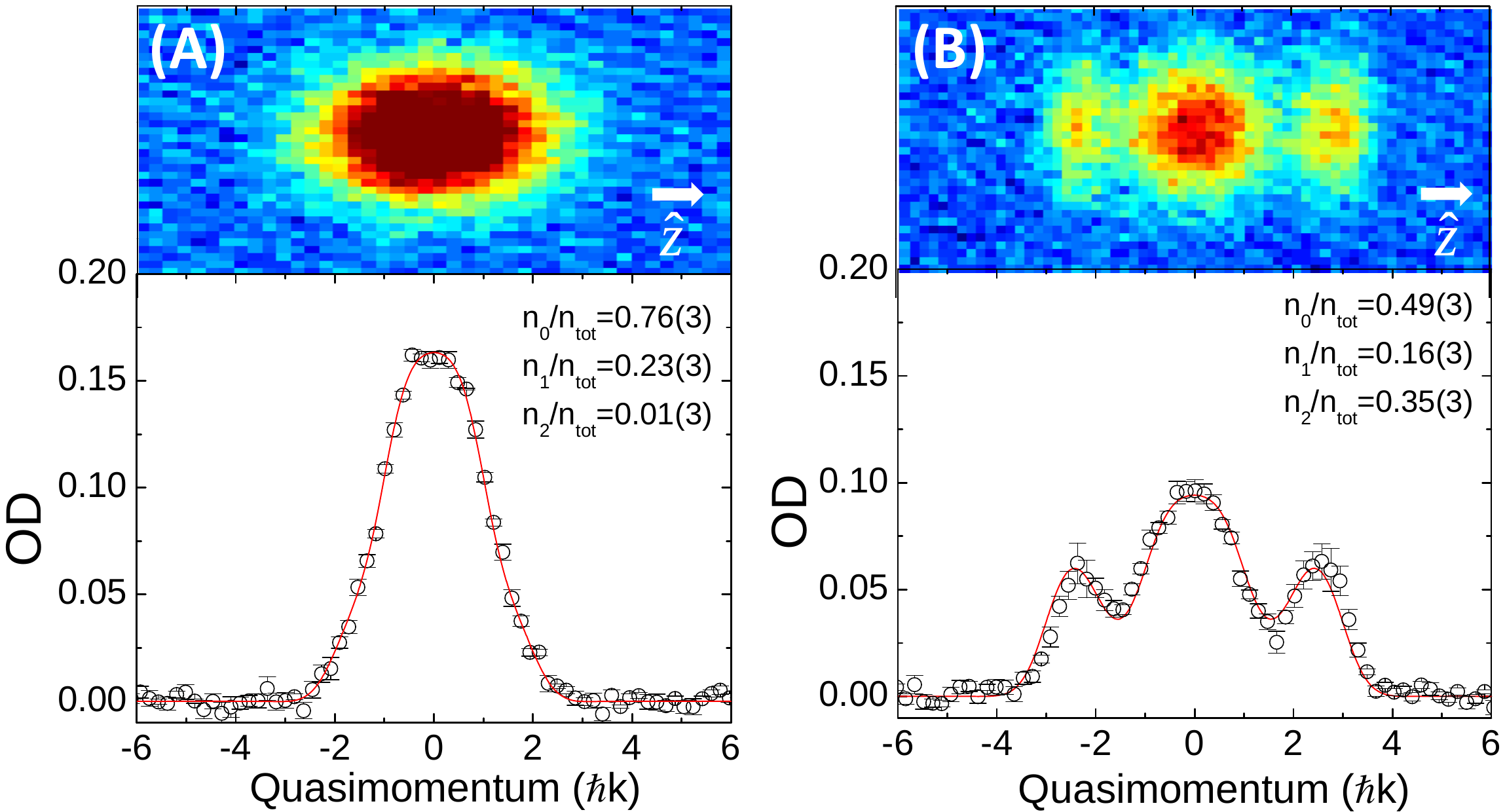}
                \caption{\textbf{Relative population of molecules in the lattice vibrational levels.}
                We measure the relative population in each lattice vibrational level using a band-mapping
                technique.  The results are shown for (A) a thermal distribution of molecules and
                 (B) a non-thermal distribution
                created by parametric heating in $\hat{z}$.  The two images use the same color scale for the optical depth
                (OD).
                The images are an average of 5 shots and 7 shots for (A) and (B), respectively,
                taken after 10 ms of free expansion. Below each image we show a trace along $\hat{z}$ that corresponds the OD averaged
                over
                the transverse direction. A fit (red line) to the
                trace, which takes into account both the size of the Brillouin zones and our imaging resolution,
                 is used to extract the relative populations, $n_v/n_{\textrm{tot}}$, in each lattice level $v$.
                 The horizontal axis corresponds to momentum in $\hat{z}$ and is marked in units of the lattice
                 momentum $\hbar k$,
                 where $k$ is the lattice wavevector. }
            \label{fig2}
        \end{figure}

We determine the population in each lattice level using an adiabatic
band-mapping technique~\cite{bandmapping1,bandmapping2}. As the
lattice potential is ramped down slowly, molecules in different
vibrational levels of the lattice are mapped onto Brillouin zones.
The measured molecule momentum distribution following this ramp is
shown in Fig.~\ref{fig2}~A for a $T=800$ nK molecular gas. The
measured fraction in $v=0$ matches well with the expected thermal
distribution. In contrast, Fig.~\ref{fig2}~B shows the
measured non-equilibrium occupation of lattice vibrational levels
following parametric heating.

We measure the bimolecular reaction rate by monitoring the loss of
trapped molecules as a function of time. To image the molecules, we
reverse our coherent transfer process to bring the molecules back to
a weakly bound state where we can detect the molecules with
time-of-flight absorption imaging~\cite{NiScience}. The molecules
are imaged after free expansion from the combined optical dipole
plus lattice trap.  From the images, we obtain the total number of
molecules and the radial cloud size.  Since we do not resolve the
individual layers of the optical lattice, we obtain an average 2D
density per layer by dividing the total number by the cross-sectional area of the cloud, and by an effective number of layers $\alpha$, as defined in the Methods section.

\begin{figure}[htbp]
        \centering
                     \includegraphics[width=1.00\textwidth]{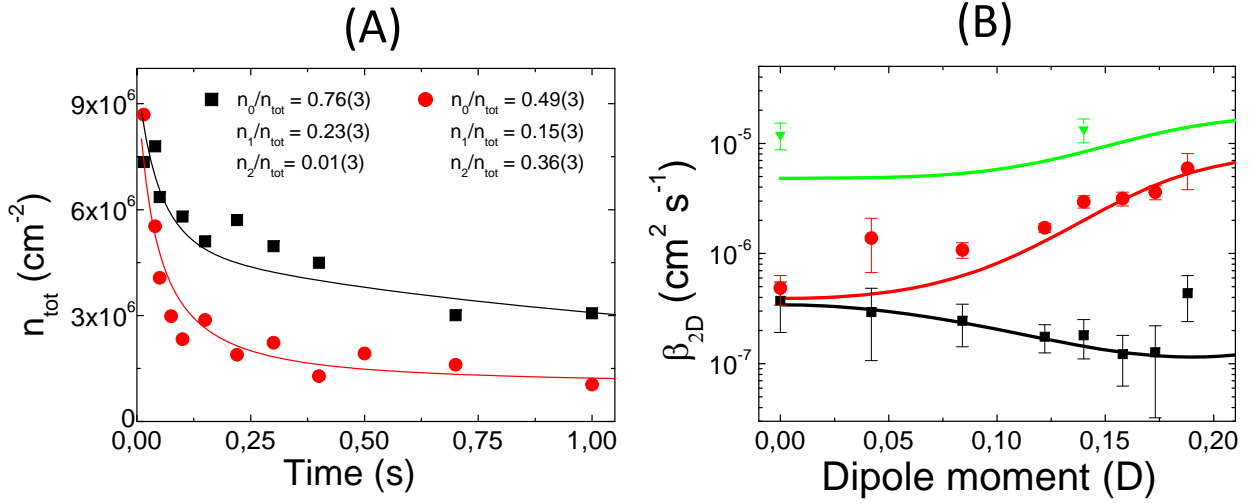}
                \caption{\textbf{Measurements of 2D loss rates and comparison with theory.} (A) A fit (solid lines) to the measured
                 loss curves, with (red circles) and without (black squares) 0.3 ms of parametric heating in $\hat{z}$, is used to
                extract the loss rate constants $\beta_{|3\rangle}$ and $\beta_{|2\rangle}$. (B) The extracted loss rate constants for collisions of molecules in the same
                lattice vibrational level (black squares) and from different lattice vibrational levels (red circles) are plotted
                for several dipole moments.  Measured loss rate constants for molecules prepared in different internal states
                are shown as green triangles. For comparison with each of these three measurements,
                we include a quantum scattering
                calculation for $\nu_z=$ 23 kHz, $T=$ 800 nK (solid lines). The potentials
                corresponding to the dominant loss channel for the three cases are shown in matching colors in Fig.~\ref{fig1}~B.}
            \label{fig3}
\end{figure}

In Fig.~\ref{fig3}~A, we show the average 2D density as a function of time. For these
data, the molecules are all prepared in the same internal state and
$|\overrightarrow{E}|$ is 4 kV/cm, which gives an induced molecular
dipole moment of 0.158 Debye (D), where 1D = 3.336 $\times 10^{-30}$
C$\cdot$m. The two data sets in Fig.~\ref{fig3}~A correspond to an
unperturbed $T=800$ nK gas (black squares) and a parametrically
heated gas (red circles). For the case where parametric heating was
used to increase population in $v>0$ levels, the data show a faster
initial loss of molecules. This suggests that the initial loss is
predominately due to interlevel collisions as described in case (2)
of Fig.~\ref{fig1}~C, while intralevel collisions (case (3) of
Fig.~\ref{fig1}~C) give a slower loss of molecules at longer times.

We fit the data using a simple model, which assumes two loss rate
constants: one for interlevel collisions, $\beta_{|2\rangle}$, and a
second one for intralevel collisions, $\beta_{|3\rangle}$ (with the subscripts referring to the adiabatic channels labeled in Fig.~\ref{fig1}~B). Here,
\begin{eqnarray}
    \frac{d{n}_{0}}{dt}=-\beta_{|3\rangle}n^{2}_{0}-\beta_{|2\rangle}n_{0}n_{1}-\beta_{|2\rangle}n_{0}n_{2},
    \nonumber \\
    \frac{d{n}_{1}}{dt}=-\beta_{|2\rangle}n_{0}n_{1}-\beta_{|3\rangle}n^{2}_{1}-\beta_{|2\rangle}n_{1}n_{2},
    \nonumber \\
    \frac{d{n}_{2}}{dt}=-\beta_{|2\rangle}n_{0}n_{2}-\beta_{|2\rangle}n_{1}n_{2}-\beta_{|3\rangle}n^{2}_{2},
\end{eqnarray}
\noindent where $n_{v}$ is the 2D density of molecules in a
particular lattice vibrational level $v$. To fit the measured time
dependence of the total 2D density, $n_{tot}(t)$, we use
$n_{\textrm{tot}}(t)=n_{0}(t)+n_{1}(t)+n_{2}(t)$. We input the
measured initial populations $n_v/n_{\textrm{tot}}$ (see
Fig.~\ref{fig2} and Methods) at $t=0$, and we fit the data to the
numerical solution of Eqn. 3. We obtain $\beta_{|3\rangle}$ and
$\beta_{|2\rangle}$ from a simultaneous fit to the two measured
$n_{\textrm{tot}}(t)$ curves shown in Fig.~\ref{fig3}~A.

By repeating this procedure for different values of
$|\overrightarrow{E}|$, we measure the chemical reaction rate
constants, $\beta_{|3\rangle}$ and $\beta_{|2\rangle}$, as a
function of the induced dipole moment. In Fig.~\ref{fig3}~B, we show
the intralevel (black squares) and interlevel (red circles) chemical
rate constants as a function of the dipole moment. Also shown as green
triangles in Fig.~\ref{fig3}~B are the results of two measurements for a 50/50
mixture of molecules in different rotational states (case (1) of
Fig.~\ref{fig1}~C). Here, we fit the loss of molecules in the ground
rotational state to the solution of
$\frac{d{n}_{\textrm{tot}}}{dt}=-\beta_{|1\rangle}
n^2_{\textrm{tot}}$ to extract a single loss rate constant.

For comparison with these measurements, we perform quantum
scattering calculations using a time-independent quantum formalism
based on spherical coordinates with cylindrical asymptotic matching
to describe the molecular collisions in quasi-2D~\cite{Quemener3}.
We use an absorbing potential at short distance to represent
chemical reactions~\cite{Quemener2, Julienne10}. This technique
showed excellent agreement with previous experimental data for KRb
bimolecular reactions in 3D~\cite{OspelScience,NiNature}. We
computed the loss rate coefficients $\beta_{v_1,v_2}$ for molecules
in different initial lattice vibrational states $v_1,v_2$, at a
collision energy of 800~nK. When the induced dipole moment is still small (0 - 0.2~D),
the measured temperature is a good approximation for the mean collision energy.
The loss rates of the different processes can be separated into fast
loss rates ($\beta_{0,1}, \beta_{0,2}, \beta_{1,2}$)$\approx
\beta_{|2\rangle}$ and slow loss rates ($\beta_{0,0}, \beta_{1,1},
\beta_{2,2}$)$\approx \beta_{|3\rangle}$. The black theoretical
curve in Fig.~\ref{fig3}~B corresponds to an average of the slow rates weighted by the
initial populations $n_{0}, n_{1}, n_{2}$. The red curve corresponds
to the same average but for the fast rates. The green curve
corresponds to the loss rate of molecules in different internal
states.

The three measured reaction rate constants shown in
Fig.~\ref{fig3}~B are consistent with the quantum scattering
calculations for the collision channels shown in matching colors in
Fig.~\ref{fig1}~B and Fig.~\ref{fig1}~C. Molecules in different
rotational states (green triangles in Fig.~\ref{fig3}~B) have the
highest rate for chemical reactions, consistent with the fact that
they can collide in channel $|1\rangle$, which corresponds to
spatially isotropic collisions with no centrifugal barrier. On the
other hand, molecules prepared in the same internal molecular state
(red circles and black squares in Fig.~\ref{fig3}~B) have suppressed
reaction rates because the lowest energy collision channel is no
longer allowed. Instead, identical molecules in different lattice
levels (red circles in Fig.~\ref{fig3}~B) react predominantly
through collisions in channel $|2\rangle$, or ``head-to-tail'',
while identical molecules in the same lattice level (black squares
in Fig.~\ref{fig3}~B) react through collisions in channel
$|3\rangle$, or ``side-by-side''. The importance of stereodynamics
on the reaction rate for polar molecules is manifest in the very
different dipole-moment dependence of the reactions rates in these
two collision channels. In particular, for the case where the
molecules are prepared both in the same internal quantum state and
in the same $v$ level, the reaction rate is suppressed even as the
dipole moment is increased.

\begin{figure}[htbp]
        \centering
                       \includegraphics[width=0.8\textwidth]{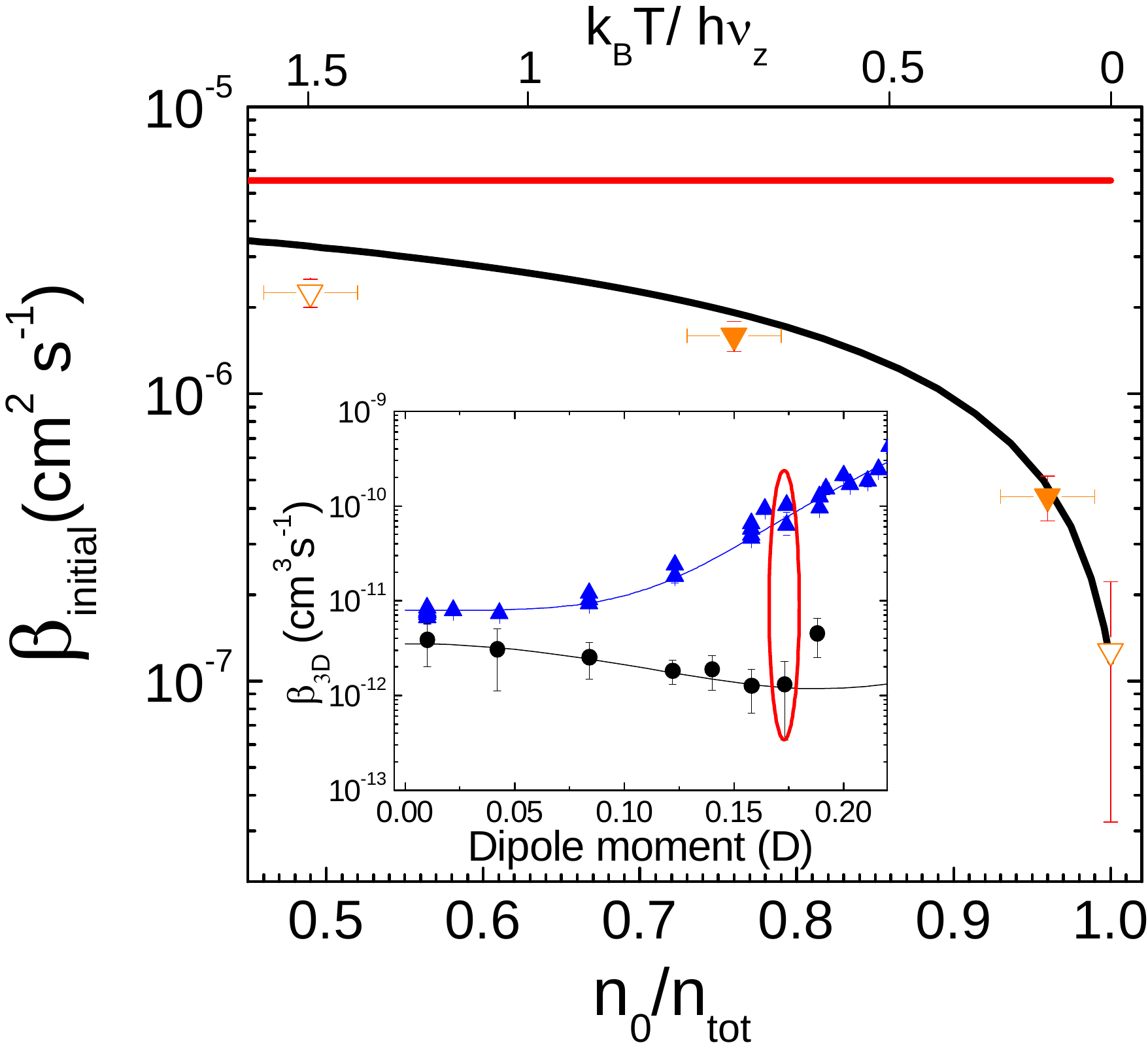}
                \caption{\textbf{Loss rates from 3D to 2D.}  The effective initial loss rate, $\beta_{\textrm{initial}}$, for polar molecules
                confined in a 2D geometry depends on the fractional population ($n_0/n_{tot}$) in the
                lowest harmonic oscillator level in $\hat{z}$, which for a gas in thermal equilibrium depends on the ratio
                $\frac{k_BT}{h\nu_z}$.
                The measured initial loss rates for a dipole moment of 0.174 D are displayed for
                two different thermal distributions (solid triangles), a non-thermal sample created
                by parametric heating (the top open triangle), and an extracted pure $\beta_{|3\rangle}$
                when the entire population is residing in the lattice ground vibrational level (the bottom
                open triangle). The experimental results agree well with
                a simple model (black curve) described in the text and Methods. The top
                line indicates the value of $\beta_{|2\rangle}$ as measured in Fig.~\ref{fig3}~B. (Inset) The intralevel loss rate for identical fermionic KRb molecules in 2D (black circles) is compared with the loss rate in 3D (blue triangles).  The 3D data for $T=300$ nK are borrowed from Ref.~\cite{NiNature}. The 2D data were taken at $T=800$ nK
                and are converted to 3D rates by multiplication with $\sqrt{\pi} a_{ho}$, where $a_{ho}$ is the harmonic oscillator length in $\hat{z}$. }
            \label{fig4}
\end{figure}

Figure~\ref{fig4} shows how the initial loss rate in a gas of
identical molecules depends on the fractional occupation of the
lowest lattice level, $n_0/n_{\textrm{tot}}$.  As
$n_0/n_{\textrm{tot}}$ increases, the calculated initial loss rate
constant for a molecular gas in thermal equilibrium (solid black
line) changes from close to $\beta_{|2\rangle}$ (the red line
indicating the measured value at 0.174 D from Fig.~\ref{fig3}~B) to
$\beta_{|3\rangle}$ (open point at $n_0/n_{\textrm{tot}}=1$). In
thermal equilibrium, the fractional occupation of the lowest
vibrational level is given by the Boltzmann distribution (see
Methods). On the top axis of Fig.~\ref{fig4}, we give the
corresponding values of the scaled temperature
$\frac{k_BT}{h\nu_z}$.  The solid triangles in Fig.~\ref{fig4}
correspond to the measured initial loss rate at different
temperatures (500 nK and 800 nK), while the open symbol at
$n_0/n_{\textrm{tot}}\approx0.5$ corresponds to the initial loss
rate for the parametrically heated, non-thermal molecular gas.

We also directly compare the suppressed chemical reaction rate in
quasi-2D to that of the 3D case in the inset to Fig.~\ref{fig4}.
Here, we compare data for a 3D geometry from Ref.~\cite{NiNature}
against the suppressed loss rate constant measured in quasi-2D.  For
the comparison, the 2D loss rate is scaled to 3D using
$\beta_{3D}=\sqrt{\pi} a_{ho}\beta_{2D}$~\cite{Petrov, Li,
Micheli3}, where $a_{ho}$ is the harmonic oscillator length in
$\hat{z}$. For a dipole moment $d$ greater than 0.1 D, the 3D loss
rate constant increases dramatically  as
$d^{6}$~\cite{NiNature,Quemener}, whereas the scaled loss rate
constant for the quasi-2D case remains close to the value at zero
electric field. At a dipole moment of 0.174 D, the measured
suppression in quasi-2D is a factor of 60.

The results shown here demonstrate how quantum stereochemistry in
the ultracold regime can be used to control reaction rates. The
capability of precisely controlling the molecular quantum states for
both the internal and external degrees of freedom is a key
ingredient for this advance. The strong suppression of reaction
rates for a gas of fermionic molecules in a quasi-2D geometry opens
the door for creating a stable ultracold gas of polar molecules and
studying the many-body physics of such a system. This approach is
particularly appealing considering that a number of bialkali polar
molecular species currently under study are expected to experience
chemical reactions at rates similar to the KRb
case~\cite{Micheli3,Hutson10,Byrd10,Meyer10}.

$\\$ $\textbf{Methods Summary}$

The traces in Fig.~\ref{fig2} were obtained by averaging the images
in the transverse direction within one rms width of the Gaussian
distribution. We fit the traces to the convolution of a series of
step functions and a Gaussian: the former describes the first three
Brillouin zones whereas the latter characterizes the effect of a
finite imaging resolution. The uncertainty in the relative
population is $3\%$, and is dominated by systematic errors arising
from the variation of the imaging resolution within the range of 1
to 2 pixels.

The average 2D density shown in Fig.~\ref{fig3}~A is obtained by
dividing the total number of molecules, $N$, by an effective area,
$4\pi\alpha\sigma_r^2$, where $\sigma_r$ is the rms cloud size in
the transverse direction and $N/\alpha$ is a number-weighted average
over the occupied lattice layers. We calculate $\alpha(t=0)$ = 23
assuming an initial discrete Gaussian distribution in $\hat{z}$ with
an rms width that we measure after transferring the molecules back
to the optical dipole trap. However, $\alpha$ increases at longer
times because of  the density dependence of the loss. For our
analysis, we use a time-averaged value $\alpha=30$ that was
determined by comparing an analysis based on a uniform layer density
to a numerical simulation of the loss in each layer. The
uncertainties for $\beta_{|3\rangle}$ and $\beta_{|2\rangle}$ are
dominated by statistical uncertainties in the fits to
$n_{\textrm{tot}}(t)$.  For the data shown in green triangles in
Fig.~\ref{fig3}~B, the two internal molecular states are
$|0,0,-4,1/2\rangle$ and $|1,1,-4,1/2\rangle$, following the
notation of Ref.~\cite{hyperfine}.

For the solid black line in Fig.~\ref{fig4}, the fractional
molecular population $f(v,T)$ in a vibrational level $v$ at
temperature $T$ is obtained from a Boltzmann distribution, and the
effective $\beta_{\textrm{initial}}$ is then calculated as
\begin{eqnarray}
   \beta_{\textrm{initial}} = \beta_{|3\rangle} \sum_v f(v,T)^2 + \beta_{|2\rangle} \sum_{v_1\neq v_2} f(v_1,T) f(v_2,T).
\end{eqnarray}

\begin{acknowledgments}
We thank P. Julienne, P. Zoller, G. Pupillo, and A. Micheli for stimulating discussions and S. Moses for technical contributions. We gratefully acknowledge financial support for this work from NIST, NSF, AFOSR-MURI, DOE, and DARPA.
\end{acknowledgments}


\end{document}